\documentclass[aps,pre,twocolumn,superscriptaddress,showpacs]{revtex4-2}
\usepackage{graphicx}
\usepackage{array}
\usepackage{amsfonts}
\usepackage{amssymb,amsmath,multirow,rotate}
\usepackage{mathrsfs}
\usepackage{booktabs}
\usepackage{threeparttable}
\usepackage{multirow}
\usepackage{epsfig}
\usepackage{threeparttable}
\usepackage{chngpage}
\usepackage{latexsym}
\usepackage{dcolumn}
\usepackage{graphics,graphicx,bm,fleqn,epic,eepic,float}
\usepackage{verbatim}
\usepackage{subfigure}
\usepackage{color}
\usepackage{xr}
\usepackage{float}
\usepackage{listings} % For including code in appendix
\usepackage{placeins} % For putting plots in the desired location
\usepackage{tikz}
\usepackage{url}
\usepackage{soul}

\usepackage{float}
\makeatletter
\let\newfloat\newfloat@ltx
\makeatother

\usepackage{algorithm}
\usepackage{algpseudocode}

\definecolor{red}{rgb}{1,0,0}

\definecolor{blue}{rgb}{0,0,1}

\begin{document}

\title{Learning to learn ecosystems from limited data - a meta-learning approach}
% \title{Learning to learn ecological systems: A meta-learning insight for system reconstruction with limited data.} % meta? or machine?
% \title{Perception of ecological system with limited data using machine learning}

\date{\today}

\author{Zheng-Meng Zhai}
\affiliation{School of Electrical, Computer and Energy Engineering, Arizona State University, Tempe, AZ 85287, USA}

%\author{Mohammadamin Moradi}
%\affiliation{School of Electrical, Computer and Energy Engineering, Arizona State University, Tempe, AZ 85287, USA}

\author{Bryan Glaz}
\affiliation{Vehicle Technology Directorate, CCDC Army Research Laboratory, 2800 Powder Mill Road, Adelphi, MD 20783-1138, USA}

\author{Mulugeta Haile}
\affiliation{Vehicle Technology Directorate, CCDC Army Research Laboratory, 6340 Rodman Road, Aberdeen Proving Ground, MD 21005-5069, USA}

\author{Ying-Cheng Lai} \email{Ying-Cheng.Lai@asu.edu}
\affiliation{School of Electrical, Computer and Energy Engineering, Arizona State University, Tempe, AZ 85287, USA}
\affiliation{Department of Physics, Arizona State University, Tempe, Arizona 85287, USA}

\begin{abstract}

A fundamental challenge in developing data-driven approaches to ecological systems for tasks such as state estimation and prediction is the paucity of the observational or measurement data. For example, modern machine-learning techniques such as deep learning or reservoir computing typically require a large quantity of data. Leveraging synthetic data from paradigmatic nonlinear but non-ecological dynamical systems, we develop a meta-learning framework with time-delayed feedforward neural networks to predict the long-term behaviors of ecological systems as characterized by their attractors. We show that the framework is capable of accurately reconstructing the ``dynamical climate'' of the ecological system with limited data. Three benchmark population models in ecology, namely the Hastings-Powell model, a three-species food chain, and the Lotka-Volterra system, are used to demonstrate the performance of the meta-learning based prediction framework. In all cases, enhanced accuracy and robustness are achieved using five to seven times less training data as compared with the corresponding machine-learning method trained solely from the ecosystem data. A number of issues affecting the prediction performance are addressed.

\end{abstract}

\maketitle

\section{Introduction} \label{sec:intro}

Recent years have witnessed a growing interest in applying machine learning to 
complex and nonlinear dynamical systems for tasks such as prediction~\cite{HSRFG:2015,LBMUCJ:2017,PLHGO:2017,LPHGBO:2017,PHGLO:2018,Carroll:2018,NS:2018,ZP:2018,GPG:2019,JL:2019,TYHNKTNNH:2019,FJZWL:2020,ZJQL:2020,KKGGM:2020,CLAC:2020,KFGL:2021a,PCGPO:2021,KLNPB:2021,FKLW:2021,KFGL:2021b,Bollt:2021,GBGB:2021,KWGHL:2023,ZKL:2023,YHBTLS:2024}, 
control~\cite{ZMKGHL:2023}, signal detection~\cite{ZMKL:2023}, and 
estimation~\cite{ZMGHL:2024}. For example, a seminal work~\cite{PHGLO:2018} 
exploited reservoir computing~\cite{Jaeger:2001,MNM:2002} to accurately predict 
the state evolution of a spatiotemporal chaotic system for about half dozen 
Lyapunov times (one Lyapunov time is the time needed for an infinitesimal error to 
grow by the factor of $e$) - a remarkable achievement considering the sensitivity
of a chaotic system to uncertainties in the initial conditions). Subsequently, 
long-term prediction of chaotic systems with infrequent state updates was 
achieved~\cite{FJZWL:2020}, and a parameter-adaptive reservoir computing was 
developed to predict critical transitions in chaotic systems based on historical 
data~\cite{KFGL:2021a,KWGHL:2023}. 
%This is a particularly challenging problem in non-autonomous dynamical systems, which had previously been deemed unsolvable for the reason that the available historical data often exhibit no sign of any potential collapse in the future, but machine learning provides a viable solution. Not only is adaptable reservoir computing able to predict critical transitions, it can predict the bifurcation diagram using data from a small number of distinct parameter values, effectively generating a digital twin of the original system~\cite{KWGHL:2023}. 

The demonstrated power of modern machine learning in solving challenging problems in 
nonlinear dynamics and complex systems naturally suggest applications to ecological 
systems that are vital to the well being of the humanity. Ecosystems in the modern 
era are nonautonomous in general due to the human-influences-caused climate change, 
and it is of paramount interest to be able to predict the future state of the 
ecosystems. However, to enable applications of machine learning to ecosystems, a 
fundamental obstacle must be overcome. Specifically, a condition under which the 
existing machine-learning methods can be applied to complex dynamical systems is 
the availability of large quantities of data for training. For physical systems 
accessible to continuous observation and measurements, this data requirement may not 
pose a significant challenge. However, for ecosystems, the available empirical 
datasets are often small and large datasets are generally notoriously difficult to 
obtain~\cite{rissman2017ecology}. A compounding factor is that ecosystems are subject 
to constant disturbances~\cite{wood2010statistical}, rendering noisy the available 
datasets. We note that, in recent years, machine learning has been applied to 
ecosystems~\cite{christin2019applications} but not for predicting the long-term 
dynamical behavior or the attractor. For example, support vector machines and random 
forests were widely used in ecological science for tasks such as classifying invasive 
plant species, identifying the disease, forecasting the effects of 
anthropogenic~\cite{drake2006modelling,cutler2007random}, estimating the hidden
differential equations~\cite{wang2011predicting,course2023state}, and 
reconstructing the ``climate'' of the entire system~\cite{wikner2023stabilizing}.
More recently, deep learning was applied in species recognition from video and audio 
analysis~\cite{tuia2022perspectives,pichler2023machine}. 

To overcome the data-shortage difficulty, we exploit 
meta-learning~\cite{finn2017model,hospedales2021meta} to predict the long-term
dynamics or the attractors of ecosystems. Meta-learning is a learning-to-learn 
paradigm that enhances the learning algorithm through experience accumulation across 
multiple episodes. Differing from the conventional machine learning approaches, a 
well-trained meta-learning framework can adapt to new tasks more swiftly and 
efficiently by leveraging its prior experience, thus reducing the necessity for 
extensive retraining and data collection. Owing to its unique features, meta-learning 
has found broad applications in fields such as computer vision~\cite{wang2019meta}, 
time series forecasting~\cite{talagala2023meta,oreshkin2021meta}, reinforcement 
learning~\cite{nagabandi2018learning}, and identification of special quantum
states~\cite{HWL:2023}. Our goal is to use meta-learning to reconstruct the ``climate'' 
of the target ecosystems, addressing the challenge of data scarcity. Specifically, we 
take advantage of a number of classical chaotic systems for training a conventional
machine-learning architecture to gain ``experience'' with complex dynamics anticipated 
to occur in ecosystems, and then update or fine-tune the machine-learning algorithm 
using the small amount of available data from the actual ecosystem. The outcome is a 
well-adapted machine-learning framework capable of predicting the complex dynamical 
behavior of ecosystems with only limited data.

What machine learning architecture is appropriate to combine the meta-learning 
algorithm for predictive modeling of ecosystems? Recurrent neural 
networks~\cite{hermans2010memory,ZMKGHL:2023,kim2023neural} such as reservoir computing 
can be a candidate since the prediction requires historical information. For 
computational efficiency, we choose time-delayed feedforward neural networks (FNNs, 
see Appendix~\ref{appendix_A})~\cite{GBGB:2021,Bollt:2021,zhang2023catch,ZMKL:2023}, a 
variant of reservoir computing, where the present and historical information of the 
time series is input into the neural network through time-delayed embedding. With 
time-delayed FNNs, the meta-learning framework becomes adept at handling the intricate 
and often nonlinear temporal dependencies typical of ecological data, thereby enabling 
it to adapt and learn rapidly from new, limited, and noisy data.

In Sec.~\ref{sec:methods}, we describe our meta-learning scheme in detail. 
In Sec.~\ref{sec:results}, successful prediction is demonstrated using three 
prototypical models: the chaotic Hastings-Powell system, a chaotic food chain, 
and the chaotic Lotka-Volterra system. Conclusions and discussion are offered in 
Sec.~\ref{sec:discussion}.

\section{Proposed meta-learning framework} \label{sec:methods}

\begin{figure*} [ht!]
\centering
\includegraphics[width=\linewidth]{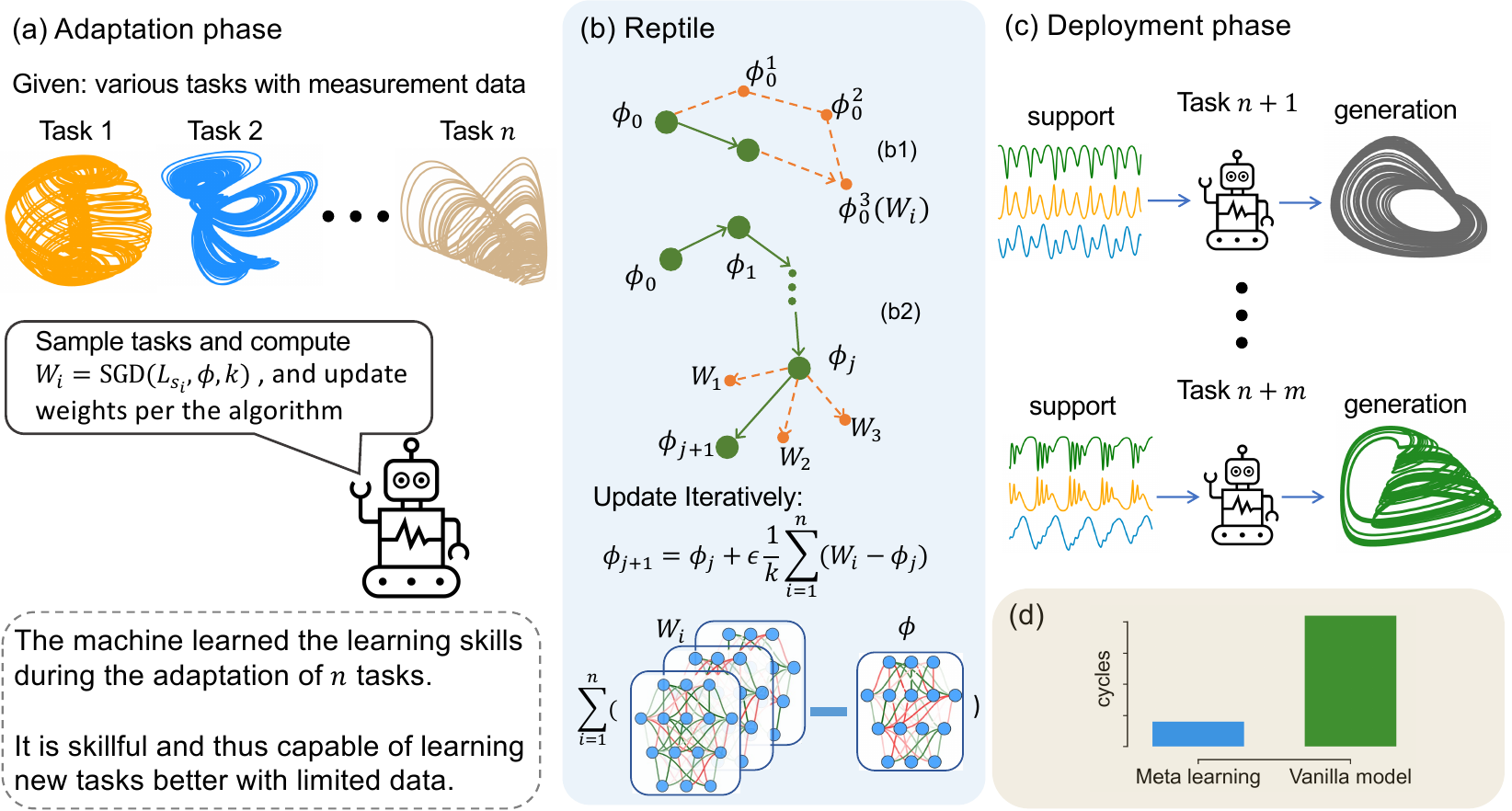} 
\caption{Illustration of the proposed meta-learning framework for reconstructing
ecosystems from limited data. (a) Adaptation phase, where the neural-network
architecture is trained on various datasets from synthetic nonlinear chaotic systems
so it learns the skill of learning and therefore can better learn the target 
ecosystem. (b) Illustration of the Reptile algorithm, a gradient-based meta-learning 
method - see text for details. (c) Deployment phase in which the trained meta-learning 
framework is applied to the target ecosystem, accomplishing the objective of predicting 
its long-term dynamics or attractor from limited time-series data. (d) An illustration 
of the comparison of the data requirements for achieving similar performance by our 
proposed meta-learning framework and standard machine-learning (vanilla model) in 
reconstructing the Hastings-Powell system.}
\label{fig:main}
\end{figure*}

The proposed meta-learning framework consists of two distinct phases: adaptation and 
deployment. In the adaptation phase, a meta-learning neural-network architecture is 
exposed to a diverse array of synthetic datasets from a number of chaotic systems,
allowing it to acquire a broad range of ``experiences,'' as illustrated in 
Fig.~\ref{fig:main}(a). This phase is crucial as it equips the neural networks with 
a versatile learning strategy, nurturing its ability to tackle new and unseen tasks
from ecosystems. For meta-learning, the Reptile algorithm, a gradient-based method, 
is standard, as shown in Fig.~\ref{fig:main}(b). Figure~\ref{fig:main}(c) illustrates 
the deployment phase, in which the well-trained meta-learning scheme is applied to a 
specific ecosystem of interest. With only limited time series data from the target
ecosystem, the scheme adeptly generates accurate long-term predictions of the 
``climate'' of the system. An issue is, as compared with a vanilla machine-learning 
scheme (not meta-learning), how much data reduction can be achieved with our 
meta-learning approach. This issue can be addressed by performing numerical experiments 
to determine the training duration required to achieve similar performance by 
meta-learning based framework and the conventional FNN model in reconstructing an 
ecosystem. Figure\ref{fig:main}(d) presents a representative result from the 
Hastings-Powell ecosystem, where the meta-learning algorithm is able to reduce the 
length of the training data approximately five times.

The core of meta-learning is the gradient-based Reptile algorithm, as shown in 
Fig.~\ref{fig:main}(b). Articulated in 2018~\cite{nichol2018first}, it has become 
a widely used method due to its simplicity and efficiency. In particular, differing 
from more complex meta-learning algorithms, Reptile requires less memory and 
computational resources, making it particularly suitable for ecosystem prediction 
from limited data. The algorithm begins by initializing the parameters. It then 
iteratively samples tasks, performs gradient descent, and updates the parameters. 
Let $\phi$ denote the parameter vector of the machine-learning architecture, $s$ 
denote a task, and SGD($L,\phi,k$) be the function performing $k$ gradient steps on 
loss $L$ starting with $\phi$ and returning the final parameter vector. The Reptile 
algorithm can be described as:
\begin{itemize}
\item Initialize $\phi$
\item For iteration = $1, 2, \ldots$, sample tasks $s_1, s_2, \ldots, s_n$
\item For $i=1, 2, \ldots, n$, compute $W_i = \text{SGD}(L_{s_i}, \phi, k)$
\item Update $\phi \gets \phi + \epsilon \frac{1}{k} \sum_{i=1}^n (W_i - \phi)$
\item Continue
\end{itemize}
Consider the loss function $L_{s_i}(\phi)$. If $\phi_t$ are the parameters at step $t$,
after $k$ steps of gradient descent, the parameters are updated as:
\begin{align}
W_i = \phi_t^{k} = \phi_t^{k-1} - \alpha \nabla_{\phi} L_{s_i}(\phi_t^{k-1}),
\end{align}
where $\alpha$ is the inner learning rate. This procedure is depicted in 
Fig~\ref{fig:main}(b1). Notably, if only one task is sampled, i.e., $s=1$, then the 
algorithm will update the parameters vector toward the direction of the green arrow 
in Fig~\ref{fig:main}(b1). Afterward, the machine trains on different tasks, as shown 
in Fig~\ref{fig:main}(b2), where the initial parameter can be updated according to
\begin{align}
\phi_{j+1} = \phi_{j} + \epsilon \frac{1}{k} \sum_{i=1}^n (W_i - \phi_j),
\end{align}
where $\epsilon$ is the outer learning rate. The outer loop procedure interpolates 
between the current initialization parameter vector $\phi$ and the task-specific 
parameter vector $W_i$, converging the parameters toward a solution close to the 
optimal solution manifold of each task $s_i$. It is worth emphasizing that the 
optimization goal of Reptile is not finding an optimal solution for a specific task 
but rather achieving a parameter vector $\phi$ that is close to the optimal solution 
manifold $\mathcal{W}_s^*$ for each task $s$. This can be formulated by minimizing 
the Euclidean distance between $\phi$ and $\mathcal{W}_s^*$:
\begin{align}
\underset{\phi}{\arg\min} \mathbb{E}_s \left[\frac{1}{2} D(\phi, \mathcal{W}_s^*)^2\right],
\end{align}
where $D(\phi, \mathcal{W}_s^*)$ denotes the distance. The gradient of this objective 
is given by
\begin{align}
\nabla_{\phi} \mathbb{E}_{s}[\frac{1}{2} D(\phi, \mathcal{W}_s^*)^2] &= \mathbb{E}_{s}[\frac{1}{2} \nabla_{\phi} D(\phi, \mathcal{W}_s^*)^2] \nonumber \\
&= \mathbb{E}_{s}[\phi-W_s^*(\phi)].
\end{align}
where $W_s^*(\phi)$ is the closest point in $\mathcal{W}_s^*$ to $\phi$. Each iteration
of Reptile samples a task $s$ and performs a stochastic gradient update:
\begin{align}
\phi &= \phi - \epsilon \nabla_{\phi} \frac{1}{2} D(\phi, \mathcal{W}^*)^2 \nonumber \\
&= \phi - \epsilon (W_s^*(\phi) - \phi) \nonumber \\
&= (1-\epsilon) \phi + \epsilon W_s^*(\phi).
\end{align}
The method ensures that the updated parameters $\phi$ are not overly specific to any 
single task, but rather effective across a range of tasks. A comparative analysis 
of Reptile with other meta-learning methods such as MAML (Model Agnostic Meta-Learning)
is presented in Appendix~\ref{appendix_B}, and the differences between meta-learning
and traditional transfer learning is discussed in Appendix~\ref{appendix_C}.

\section{Results} \label{sec:results}

We present forecasting results for three ecosystems: the three-dimensional chaotic 
Hastings-Powell system~\cite{hastings1991chaos}, a three-species food chain 
system~\cite{mccann1994nonlinear}, and the Lotka-Volterra system~\cite{vano2006chaos}
with three species. The hypothesis is that the observational data from each system 
is quite limited (to be quantified below), so it is necessary to invoke meta-learning 
by first training the neural network using synthetic time series from the 
computational models of a number of prototypical chaotic systems. Since the target 
ecosystems are three-dimensional, the chosen chaotic systems should have the same 
dimension. We use 27 such chaotic systems, as described in Appendix~\ref{appendix_D}.   
More specifically, during the adaptation phase, the neural-network architecture
is trained and the values of the hyperparameters are determined with time-series data 
from the 27 synthetic systems. In the deployment phase, further training with
appropriate adjustments to the hyperparameter values is done with the limited 
data from the target ecosystem, followed by prediction of its long-term dynamics. 
It is worth noting that the continuous adjustments and fine-tuning of the parameters
is the key feature that distinguishes meta-learning from transfer leaning, as further
explained in Appendix~\ref{appendix_C}. To demonstrate the superiority of
meta-learning to conventional machine learning, we train the {\em same} neural-network 
architecture but using time series from the ecosystems only without any pre-training - 
the so-called vanilla or benchmark machine-learning scheme. For the vanilla scheme, 
typically much larger datasets are required to achieve comparable prediction 
performance by meta-learning.

Unless otherwise stated, the following computational settings are used. Given a target 
system, time series are generated numerically by integrating the synthetic system 
models with the time step $dt=0.01$. The initial states of both the dynamical process 
and the neural network are randomly chosen from a uniform distribution. An initial 
segment of the time series is removed to ensure that the trajectory has reached the 
attractor. The training and testing data are obtained by sampling the time series at 
the interval $\Delta_s$. Specifically, for the chaotic Hastings-Powell, food-chain, 
and Lotka-Volterra systems, we set $\Delta_s=60dt=0.6$, $0.5$, and $0.2$, corresponding 
to approximately $1/77$, $1/83$, and $1/71$ cycles of oscillation, respectively. The 
time series data are preprocessed by using min-max normalization so that they lie in 
the unit interval [0,1]. Considering the omnipresence of noise, we add Gaussian noise 
of amplitude $\sigma=0.003$ to the normalized data. The training and predicting lengths 
of systems are set as 20,000 and 50,000, respectively. For the time-delayed FNN, 
the embedding dimension is 1,000, so the dimension of the input vector is 3,000 
(three-dimensional systems). The neural network comprises three hidden layers with 
the respective sizes [1024,512,128], and its output layer has the size of three (for 
three-dimensional target systems). The batch size is set to be 128. In meta-learning,
we specify 20 inner iterations ($I_i$) and 30 outer iterations ($I_o$), with the inner 
and outer learning rates of $\alpha=10^{-3}$ and $\epsilon=1$, respectively. We apply 
Bayesian optimization to systematically determine the optimal hyperparameters and 
and test the effects of the hyperparameters on the performance. Simulations are run 
using Python on two desktop computers with 32 CPU cores, 128 GB memory, and one RTX 
4000 NVIDIA GPU.

To make the presentation succinct, in the main text we focus on the results from the 
chaotic Hastings-Powell system with a brief mentioning of the summarizing results
for the three-species food chain and the chaotic Lotka-Volterra systems. The detailed
results from the latter two systems are presented in Appendix~\ref{appendix_E}. It is 
worth noting that the chaotic Hastings-Powell system is a seminal model in population 
dynamics. It describes the feeding relationships in a food chain from prey to 
predators~\cite{hastings1991chaos} and has inspired numerous variations and studies. 
The three-species food chain system~\cite{mccann1994nonlinear} is in fact one variant
of the Hastings-Powell system, exhibiting a wide range of behaviors due to the 
incorporation of additional factors and bioenergetically derived parameters. There 
were also substantial works based on the original chaotic Hastings-Powell 
model~\cite{hastings2018transient,proulx2005network,kot2001elements,terborgh2013trophic}, 
making it a benchmark and prototypical model in theoretical ecology. 

\subsection{Forecasting the chaotic Hastings-Powell system}

\begin{figure*} [ht!]
\centering
\includegraphics[width=\linewidth]{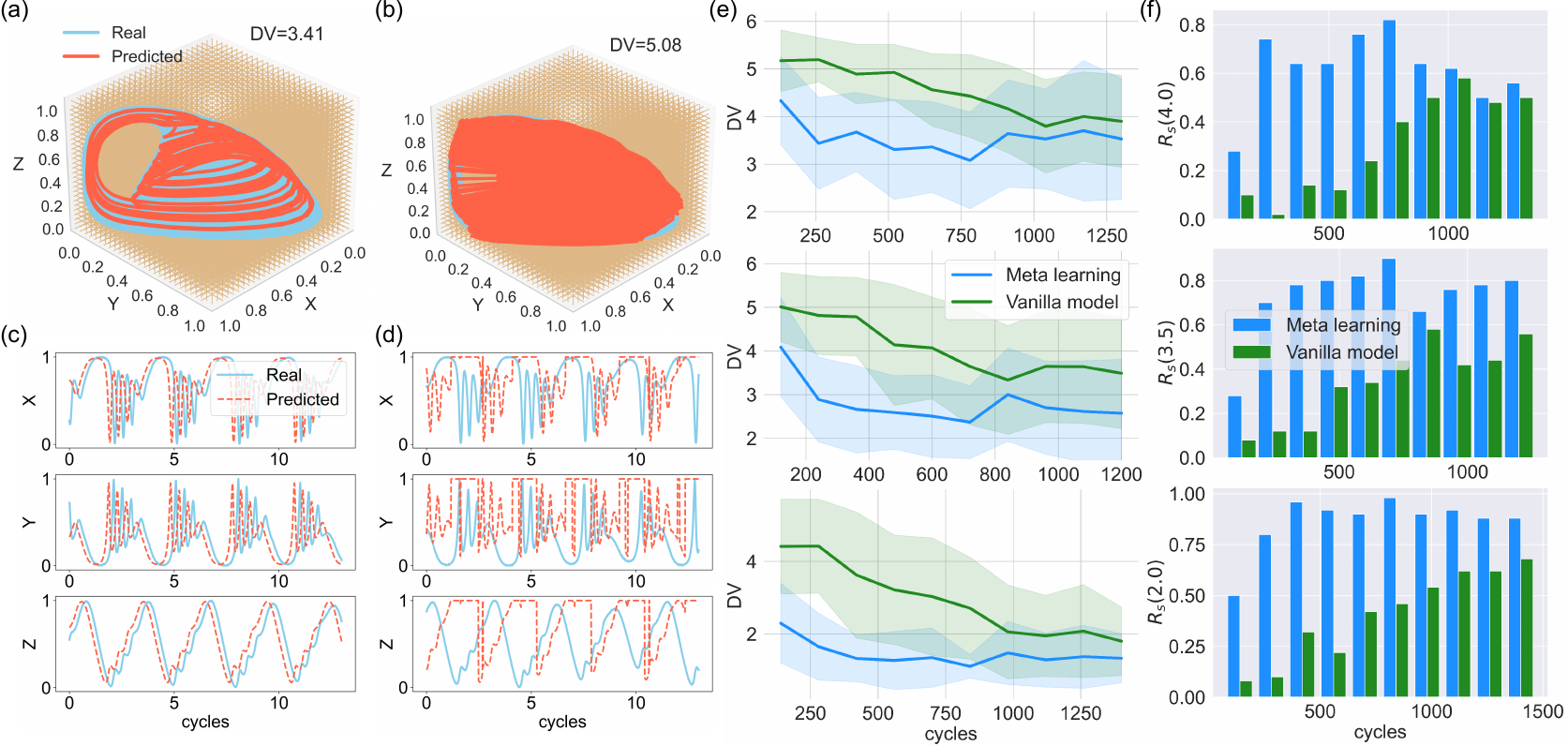} 
\caption{Main result: long-term ecosystem prediction by the meta-learning and vanilla 
frameworks. (a,b) Attractor reconstruction by the two frameworks. (c,d) Intercepted 
snippets of the three time series of the ground truth and prediction by the two 
frameworks. (e) DV versus the training length for the meta-learning and vanilla 
frameworks. (f) Stability indicator of prediction ($R_s({\rm DV}_c)$) versus the 
training length for the meta-learning and vanilla frameworks. The upper, middle, and 
lower panels in (e) and (f) are from the chaotic Hastings-Powell, food chain, and 
Lotka-Volterra systems, respectively. To reduce the statistical fluctuations, the DVs,
their shaded variabilities and the $R_s({\rm DV}_c)$ values are calculated from an 
ensemble of 50 independently trained neural machines.}
\label{fig:dv_comparison}
\end{figure*}

The chaotic Hastings-Powell system~\cite{hastings1991chaos} has three dynamical 
variables, corresponding to the resource, consumer, and predator abundances, 
respectively. The system is described by the following set of differential 
equations~\cite{vissio2023weather}:
\begin{align}\label{eq:hastings}
\frac{dV}{dt} &= V (1-V) - \frac{a_1 V H}{b_1 V + 1}, \nonumber \\
\frac{dH}{dt} &= \frac{a_1 V H}{b_1 V + 1} - \frac{a_2 H P}{b_2 H + 1} - d_1 H, \\
\frac{d P}{dt} &= \frac{a_2 H P}{b_2 H + 1} - d_2 P, \nonumber 
\end{align}
where $V$, $H$, and $P$ are the biomass of the vegetation (or resource), herbivore 
(or consumer), and predator species, respectively. Alternatively, they can also 
represent vegetation, host, and parasitoid. The parameters $a_1$, $a_2$, $b_1$, $b_2$, 
$d_1$, and $d_2$ are chosen to be biologically reasonable~\cite{hastings1991chaos} 
as $5, 0.1, 3, 2, 0.4$ and $0.01$, respectively, which contain the information about 
the growth rate, the carrying capacity of the vegetation, etc.

To characterize the performance of long-term prediction of the attractor, we use 
two measures: deviation value (DV) and prediction stability, where the former 
describes the distance between the ground truth and predicted attractors and the 
latter (denoted as $R_s({\rm DV}_c)$) is the probability that meta-learning 
generates stable dynamical evolution of the target ecosystem in a fixed time window. 
The definition of the three-dimensional DV here is extended from its two-dimensional 
version~\cite{ZKL:2023}. To calculate the DV value, we place a uniform lattice in the 
three-dimensional phase space with the cell size $\Delta=0.04$ and count the number of 
trajectory points in each cell for both the true and predicted attractors in a fixed 
time interval. The DV is given by
\begin{align}\label{eq:DV}
{\rm DV} \equiv \sum_{i=1}^{m_x}\sum_{j=1}^{m_y}\sum_{k=1}^{m_z} \sqrt{(f_{i,j,k} - \hat{f}_{i,j,k})^2},
\end{align}
where $m_x$, $m_y$, and $m_z$ are the total numbers of cells in the $x$, $y$, and 
$z$ directions, respectively, $f_{i,j,k}$ and $\hat{f}_{i,j,k}$ are the frequencies 
of visit to the cell $(i,j,k)$ by the true and predicted trajectories, respectively. 
When the predicted trajectory leaves the square, we count them as if they belonged 
to the cells at the boundary where the true trajectory never visits. To obtain the 
prediction stability, we perform the experiment $n$ times and calculate the probability 
that the DV is below a predefined stable threshold, which is given by
\begin{align}
R_s({\rm DV}_c) = \frac{1}{n} \sum_{i=1}^n [{\rm DV} < {\rm DV}_c], 
\end{align}
where ${\rm DV}_c$ is the DV threshold, $n$ is the number of iterations and $[\cdot]=1$ 
if the statement inside is true and zero otherwise.

Figure~\ref{fig:dv_comparison} presents the comparative forecasting results, where
Figs.~\ref{fig:dv_comparison}(a) and \ref{fig:dv_comparison}(b) show the ground 
truth and the predicted attractors in the three-dimensional space by the meta-learning 
and vanilla frameworks, respectively. It can be seen that the attractor predicted 
by meta-learning has a lower DV, indicating that the predicted attractor is closer 
to the ground truth. Figures~\ref{fig:dv_comparison}(c) and \ref{fig:dv_comparison}(d) 
display some representative time-series segments of the predicted and true attractors
from the meta-learning and vanilla frameworks, respectively, using the same training 
data from the ecosystem. Apparently, the vanilla framework fails to predict the 
attractor correctly. Figures~\ref{fig:dv_comparison}(e) and \ref{fig:dv_comparison}(f) 
show, respectively, the DV and prediction stability values versus the length of the 
training data from the three ecosystems. The meta-learning framework consistently 
yields a lower testing DV and higher prediction stability compared to those from 
the vanilla framework, indicating that meta-learning not only predicts more accurately 
the long-term dynamics on the attractor but the results are also more stable and 
reliable. These advantages are particularly pronounced with shorter training lengths. 
In terms of the DV indicator, the vanilla framework requires approximately 5 to 7 
times the amount of training data in meta-learning to achieve a similar level of 
performance. In terms of the prediction stability, the vanilla framework fails to 
match the performance of meta-learning, regardless of the training length.

In meta-learning, determining the optimal values of the hyperparameters is key to 
achieving reliable and accurate prediction results, which is done through standard 
Bayesian optimization. The procedure and the role of the optimal hyperparameter 
values are discussed in Appendix~\ref{appendix_F}.

\subsection{Robustness against noise}

\begin{figure} [ht!]
\centering
\includegraphics[width=\linewidth]{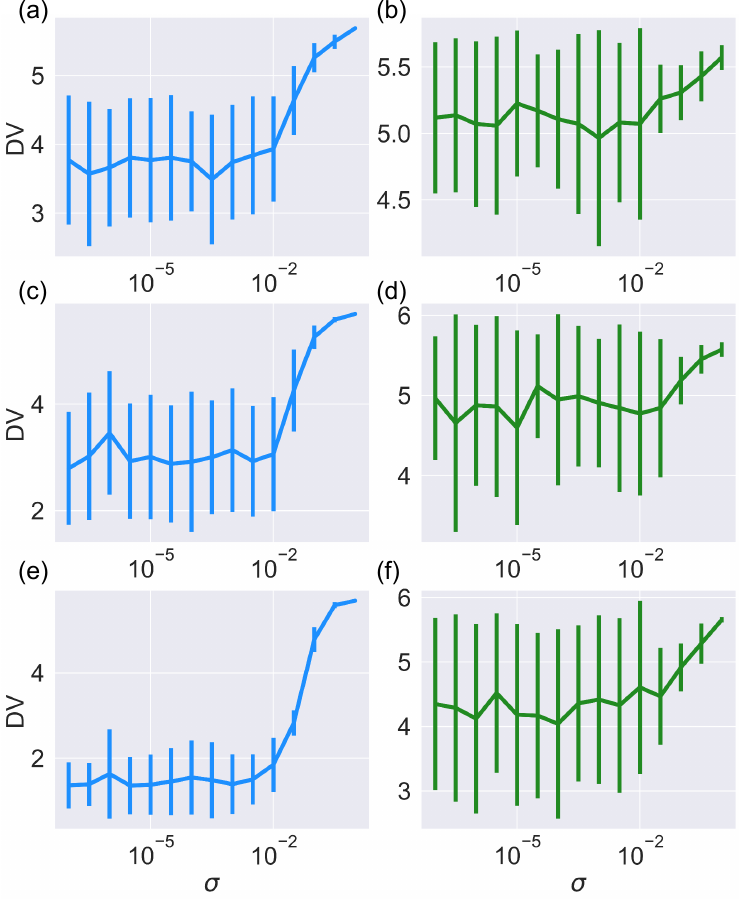} 
\caption{Robustness of ecosystem forecasting against environmental noise. The top, 
middle, and bottom rows are attractor-prediction results for the chaotic 
Hastings-Powell, food chain, and Lotka-Volterra systems, respectively. The left and 
right columns are results from the meta-learning and vanilla frameworks, respectively. 
The same training lengths are used in all cases. The average prediction DV and the 
error bar are obtained from 50 independent training and testing runs. For meta-learning, 
the DVs remain small and approximately constant when the noise amplitude is below 
$10^{-2}$. Overall, the prediction results from meta-learning are more accurate and 
robust against noise than the vanilla framework.}
\label{fig:noise_test}
\end{figure}

It is useful to study how ecosystem forecasting is degraded by environmental 
noise for the meta-learning and vanilla frameworks. We consider normally distributed 
stochastic processes of zero mean and standard deviation $\sigma$ to simulate the 
noise, which perturbs the time series point $x$ to $x+\xi$ after normalization. 
Figure~\ref{fig:noise_test} illustrates the effects of the noise on predicting the 
three ecosystems via the meta-learning and vanilla frameworks, where 50 independent 
realizations are used to calculate the average DV and the associated uncertainty. 
The results indicate that, for small noise (e.g., amplitude less than $10^{-2}$), 
robust prediction performance as characterized by relatively low DVs can be achieved 
by meta-learning. In contrast, the DVs from the vanilla framework are higher. The 
results in Fig.~\ref{fig:noise_test} suggest that meta-learning is capable of 
yielding more accurate and robust prediction results.  

\begin{figure} [ht!]
\centering
\includegraphics[width=\linewidth]{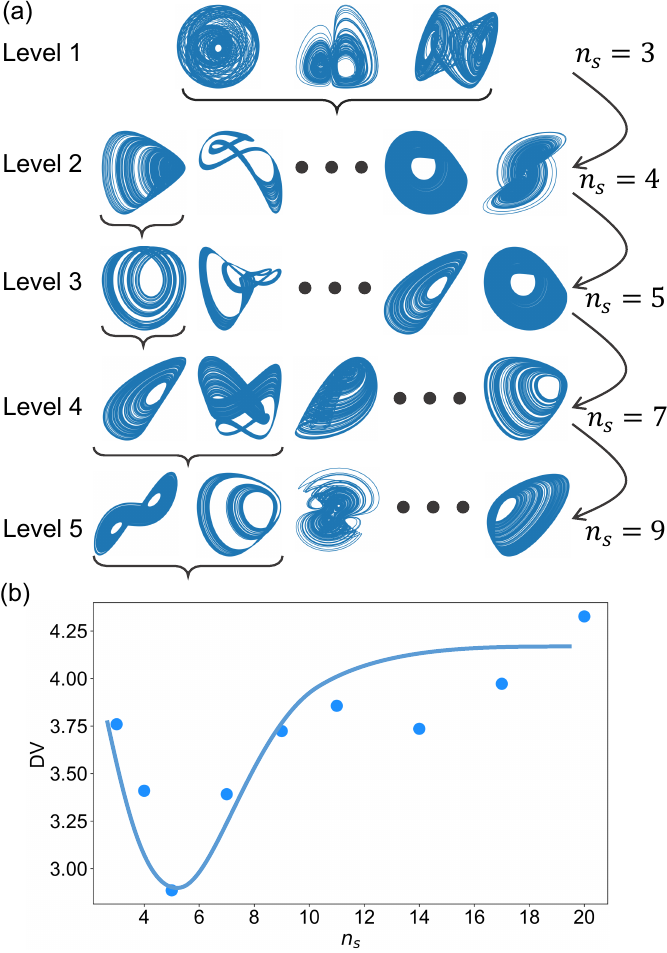} 
\caption{Selecting the synthetic chaotic systems for the adaptation phase of 
meta-learning. (a) Illustration of greedy algorithm. Stated with the three systems 
in the sampled systems pool, one or several systems is (are) selected which lead to
the best improvement in performance. (b) Ensemble averaged DV (with 50 independent 
realizations) versus the number of sampled systems pool $n_s$. As $n_s$ increases, 
the average DV decreases rapidly but later increases again.}
\label{fig:greedy_algorithm}
\end{figure}

\subsection{Optimal selection of synthetic systems for meta-learning}

As described in Sec.~\ref{sec:methods}, the superiority of the meta-learning framework 
lies in gathering experience during the adaptation phase. However, indiscriminately 
utilizing different alternative chaotic systems as adaptation tasks in general does 
not lead to desired performances, and even worse, can destroy the training performance 
owing to the diversity of such systems. This raises the question of how to choose 
the proper adaptation systems for meta learning. To address this question, we employ
the greedy algorithm to choose the optimal synthetic chaotic systems and use the 
chaotic Hastings-Powell system as a testing case for this algorithm. The test is 
performed, as follows. At each iteration, we perform a testing loop that involves 
adding a candidate system to the existing pool of chosen systems and monitoring the 
corresponding decrease in the resulting DV. Afterward, we remove this system and 
test another candidate system. After looping over all the candidate systems, we select 
one or several systems that lead to the maximal reduction of the average DV value 
calculated from 50 independent runs. Once a system has been selected, it will become 
a member of the chosen system pool in the iterative process that follows. 

Figure~\ref{fig:greedy_algorithm}(a) depicts this selection process, where the initial
sampled system pool is $n_s=3$. Looping at level one informs us that a new system 
should be added to the sampled system pool, so at Level two the pool size becomes 
$n_s=4$. Repeating this process iteratively, we collect the ensemble-averaged DV 
and the corresponding sampled system pool $n_s$, as shown in 
Fig.~\ref{fig:greedy_algorithm}(b). We observe that the average DV decreases rapidly 
as the number of systems increases from one to twenty but begins to increase again 
when more systems are included. Consequently, guided by the greedy algorithm, we 
select the five most effective systems: Aizawa, Bouali, Chua, Sprott third, and 
Sprott fourteen (See Appendix~\ref{appendix_D} for a detailed description
of these systems). It is worth noting that we do not expect the greedy algorithm to 
produce globally optimal solutions in the space of all possible chaotic systems. It 
might miss useful systems, each alone would not reduce the DV but their combination 
would. Considering that that this feature selection process is NP-hard, finding some 
locally optimal solutions is reasonable. This also implies that, while the performance 
of meta-learning is remarkable, there is ample room for improvement.

\section{Discussion} \label{sec:discussion}

Exploiting machine learning to predict the behaviors of dynamical systems has 
attracted extensive research in recent years, and it has been demonstrated that 
modern machine learning can solve challenging problems in complex and nonlinear
dynamics that were previously deemed unsolvable. However, machine-learning algorithms 
often require extensive data for training, and this presents a significant challenge
for ecosystems. Indeed, the observational datasets for ecosystems, especially those 
described by the population dynamics, are often small, preventing a straightforward
and direct application of machine learning to these systems. 

This work develops an ``indirect,'' meta-learning framework for forecasting the 
long-term dynamical behaviors of chaotic ecosystems through a faithful reconstruction 
of the attractor using only limited data. Given a chaotic ecosystem of interest, the 
idea is to use a large number of alternative chaotic systems of the same dimension, 
which can be simulated to generate massive training data for a suitable 
machine-learning scheme such as the time-delayed feedforward neural-network 
architecture. The neural networks are trained using the synthetic data first, and 
are then ``fine-tuned'' with the data from the actual target ecosystem. As a result 
of the pre-training or first-stage training for adaptation, the neural machine is 
sufficiently exposed to the climate of the dynamical evolution of characteristically 
similar systems, which can then be readily adapted to the ecosystem. Specifically, we 
employed Reptile as the meta-learning algorithm. During the adaptation phase, the 
algorithm begins by gaining ``experience'' from learning a synthetic chaotic system. 
This process continues with data from different non-ecological chaotic systems until the 
machine is well-trained, experienced, and able to learn new tasks with limited data. 
In the deployment phase, the neural machine is further trained using the limited data 
from the target ecosystem - the second-stage training. we emphasize that 
the first-stage training uses massive data from a large number of model chaotic 
systems, and the second-stage training is done with limited data from the target 
ecosystem. After the second-stage training, the neural machine is capable of 
generating the correct attractor of the ecosystem, realizing accurate and reliable 
forecasting of its long-term dynamics. It is worth noting that, while the systems
used in the adaptation phase of the training have characteristically similar 
attractors, their detailed dynamical evolution can still be quite distinct. 
It can then be intuitively anticipated that the meta-learning framework will
not have any advantage over the vanilla model with respect to short-term prediction
of the state evolution (see Appendix~\ref{appendix_G}). 

One feature of our meta-learning framework is the integration of time-delayed 
feedforward neural networks for processing sequential data. It incorporates the 
concept of time delays into the conventional FNN architecture so as to take the 
advantage of the present and historical information in the time series. We have 
tested the meta-learning framework on three benchmark ecosystems. In all cases, more 
accurate and robust prediction was achieved by the meta-learning based framework, 
whereas the vanilla model requires $5-7$ times the training data to achieve a similar 
performance. Issues such as the effect of noise and the number of synthetic systems 
used in the adaptation phase of the training were addressed.

In general, meta-learning is a powerful machine-learning tool for solving prediction
and classification problems in situations where the available data amount is small.
A recent example is detecting quantum scars in systems with chaotic classical dynamics. 
In particular, in a closed quantum system in the semiclassical regime where the 
particle wavelength is much smaller than the system size, a vastly large number of 
eigenstates are permitted, among which are those whose wavefunctions are not uniformly 
distributed in the physical space but instead concentrate on some classical periodic 
orbits of low periods. The emergence of such scarring eigenstates is counterintuitive, 
as the classical trajectories are uniform due to 
ergodicity~\cite{mcdonald1979spectrum,heller1984bound}. In the field of quantum chaos, 
traditionally identifying quantum scarring states was done in a ``manual'' way through 
a visual check of a large number (e.g., $10^4$) of eigenstates~\cite{XHLG:2013}. This 
was challenging as the percentage of scarring states is typically small - less than 
$10\%$ of all the eigenstates. A recent work demonstrated that meta-learning can be 
powerful for accurately detecting quantum scars in a fully automated and efficient 
way~\cite{HWL:2023}, where a standard large dataset called Omniglot from the field 
of image classification was used for training in the adaptation phase. 

Our meta-learning framework incorporating time-delayed FNNs possesses a high level 
of sensitivity to the temporal variations in the data, making it potentially feasible 
for extension beyond ecosystems to challenging problems such as epidemic spread 
prediction and traffic forecasting, where effective data collection is often a hurdle. 
There is also room for enhancing the performance of the framework. For example, in the 
present work, we employed the greedy algorithm for selecting synthetic chaotic systems 
for the adaptation phase of the training and implement meta-learning using the Reptile 
algorithm. Alternative algorithms can be exploited to achieve better performance.

\section*{Data and Code Availability}

Data and code are available upon request and will be available on GitHub.

\section*{Acknowledgment}

This work was supported by the Air Force Office of Scientific Research under 
Grant No.~FA9550-21-1-0438 and by the Army Research Office.

\appendix

\section{Time-delayed Feedforward neural networks} \label{appendix_A}

\begin{figure} [ht!]
\centering
\includegraphics[width=\linewidth]{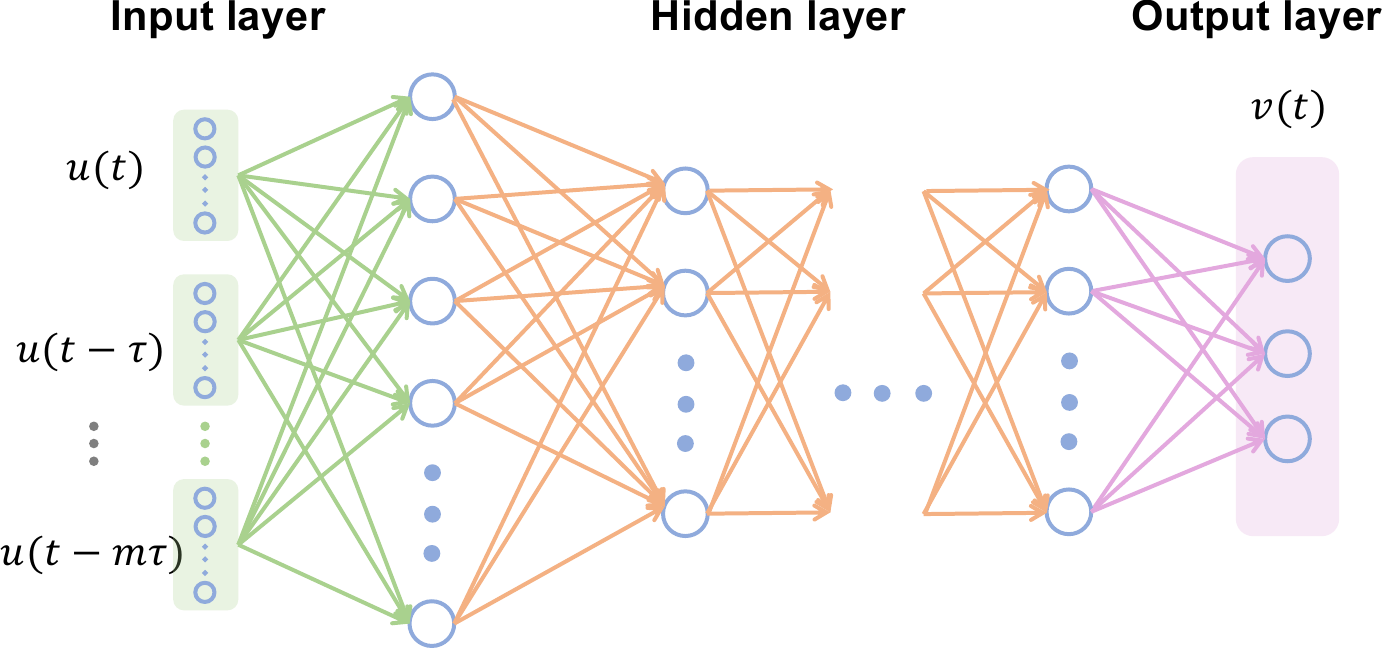} 
\caption{Architecture of a time-delayed feedforward neural network. There are three 
main components: the input layer, a number of hidden layers, and the output layer, 
where $u(t), u(t-\tau), \cdots, u(t-m\tau)$ are the present and historical signal 
with $\tau$ being the time delay and $m$ being the embedding dimension. The output is 
$v(t)$.}
\label{fig:mlp}
\end{figure}

FNNs have a straightforward neural-network architecture. Compared with recurrent neural 
networks, information in an FNN moves in only one direction, i.e., the neurons only 
process the information forward through the chain of connections from the input to the 
output. An FNN consists of layers of neurons, where each neuron in each layer has direct 
connections to the neurons in the subsequent layer. These connections are characterized 
by weights to be adjusted during the training process. A sufficiently large FNN can be 
deemed as a universal nonlinear approximator, which can capture the hidden complex 
relationship between the input and output data.

We use time-delayed FNNs~\cite{GBGB:2021,ZMKL:2023}, as illustrated in Fig.~\ref{fig:mlp},
which incorporates time delays into the original FNN architecture so as to instill
certain memory function into the original FNN. A time-delayed FNN provides the present 
and historical information of the time series, making it suitable for processing 
sequential data. Specifically, to embed the memory into a FNN, we use the present signal 
$u(t)$ and the historical signals $u(t-\tau), \cdots, u(t-m\tau)$ as the inputs, where
$\tau$ is the delayed time and $m$ is the system dimension.

The loss function is the mean squared error (MSE), which quantifies the difference 
between the predicted output and the true output:
\begin{align} \label{eq:mse}
{\rm{MSE}} = \frac{1}{N_m} \sum_{i=1}^{N_m} (\hat{y}_i - y_i) ^ 2,
\end{align}
where $N_m$ is the number of samples, $\hat{y}$ and $y$ are the predicted and true 
outputs, respectively. We utilize the ReLU (Rectified Linear Unit) activation function 
defined as $f(x)=\max(0, x)$. It introduces nonlinearity into the network, enabling 
it to learn complex patterns and relationships in the data, and it is computationally 
efficient. In addition, for optimizing the weights in a time-delayed FNN, we employ 
the stochastic gradient descent (SGD) algorithm, specifically the Adam (Adaptive Moment 
Estimation) variant, which computes the adaptive learning rates for each parameter by 
storing exponentially decaying average of the past squared gradients. The adaptive 
ability makes it suitable and effective for handling non-convex optimization problems 
commonly arisen in neural networks.

\section{Reptile algorithm} \label{appendix_B}

Here we provide a mathematical description of the Reptile algorithm~\cite{nichol2018first}.
We compare it with MAML~\cite{finn2017model} and FOMAML (First Order MAML), focusing on 
the case of two gradient steps ($k=2$) in the SGD process. Similar to the definitions in 
the main text, we denote $L^0$ and $L^1$ as the losses for different batches of data. Let 
$\phi$ be the parameter vector and $\alpha$ be the step size. For simplicity, we annotate 
$g_j^i=\nabla_{\phi} L^i (\phi_j)$ and $h_j^i=\nabla^2_{\phi} L^i (\phi_j)$. The 
parameter updates after two SGD steps are
\begin{align}
\phi_0 &= \phi \\
\phi_1 &= \phi_0 - \alpha \nabla_{\phi} L^0(\phi_0) = \phi_0 - \alpha g_0^0 \\
\phi_2 &= \phi_1 - \alpha \nabla_{\phi} L^1(\phi_1) = \phi_0 - \alpha g_0^0 - \alpha g_1^1 
\end{align}
The gradient of Reptile, MAML, and FOMAML are defined as:
\begin{align} 
\label{eq:gradient1}
g_{\rm Reptile} &= (\phi_0-\phi_2) / \alpha = g_0^0 + g_1^1, \\
\label{eq:gradient2}
g_{\rm MAML} &= \nabla_{\phi_0} L^1 (\phi_1) \nonumber \\
&= \nabla_{\phi_1} L^1 (\phi_1)(I-\alpha \nabla_{\phi}^2l^0(\phi_0)) \nonumber \\
&= g_1^1 - \alpha h_0^0 g_1^1, \\
\label{eq:gradient3}
g_{\rm FOMAML} &= \nabla_{\phi_1} L^1(\phi_1) = g_1^1.
\end{align}
Applying Taylor expansion to $g_1^1$ at $\phi_0$, we get:
\begin{align}\label{eq:taylor}
g_1^1 &= \nabla_{\phi_0} L^1(\phi_1) \nonumber \\
&= \nabla_{\phi_0} L^1(\phi_0) + \nabla_{\phi_0}^2 L^1 (\phi_0) (\phi_1-\phi_0) + O(\alpha^2) \nonumber \\
&=g_0^1 -\alpha h_0^1 g_0^0 + O(\alpha^2)
\end{align}
Substituting this into the gradients of Reptile, MAML, and FOMAML in~\ref{eq:gradient1} 
to \ref{eq:gradient3}, we get 
\begin{align}
g_{\rm Reptile} &= g_0^0 + g_0^1 - \alpha h_0^1 g_0^0 + O(\alpha^2) \\
g_{\rm MAML} &= g_0^1 - \alpha h_0^1 g_0^0 - \alpha h_0^0 g_0^1 + O(\alpha^2) \\
g_{\rm FOMAML} &= g_0^1 - \alpha h_0^1 g_0^0 + O(\alpha^2)
\end{align}
During the training, we use batches 0 and 1 to calculate losses $L^0$ and $L^1$, 
respectively. Let $\mathbb{E}_{s, 0, 1}$ be the expectation value over the two batches 
in the task $s$ and let 
\begin{align} \nonumber
	{\rm AvgGrad} = \mathbb{E}_{s,0,1}[g_0^0]=\mathbb{E}_{s,0,1}[g_0^1] 
\end{align}
be the gradient of the expected loss. The negative direction of AvgGrad brings $\phi$ 
towards the minimum of the ``joint training'' problem so as to achieve better task 
performance. Let 
\begin{align} \nonumber
	\mbox{AvgGradInner} &= \mathbb{E}_{s, 0, 1}[h_0^1 g_0^0]=\frac{1}{2}\mathbb{E}_{s, 0, 1}[h_0^1 g_0^0 + h_0^0 g_0^1] \\ \nonumber
	&= \frac{1}{2}\mathbb{E}_{s, 0, 1}[\nabla_{\phi}(g_0^0g_0^1)]. 
\end{align}
The negative direction of AvgGradInner thus increases the inner product between gradients 
of different batches for the given task so as to yield better generalization.

The gradients for the meta-learning algorithms, aiming for optimal task performance and 
generalization, can be summarized as:
\begin{align}
\mathbb{E}_{s, 0, 1}[g_{\rm Reptile}] &= 2 {\rm AvgGrad} - \alpha \cdot {\rm AvgGradInner} + O(\alpha^2), \\
\mathbb{E}_{s, 0, 1}[g_{\rm MAML}] &=  {\rm AvgGrad} - 2 \alpha \cdot {\rm AvgGradInner} + O(\alpha^2), \\
\mathbb{E}_{s, 0, 1}[g_{\rm FOMAML}] &=  {\rm AvgGrad} - \alpha \cdot {\rm AvgGradInner} + O(\alpha^2),
\end{align}
where each gradient expression first leads to the minimum of the expected loss 
over tasks, and then the higher order AvgGradInner term enables fast learning.

\section{Meta-learning versus transfer learning} \label{appendix_C}

\begin{figure} [ht!]
\centering
\includegraphics[width=\linewidth]{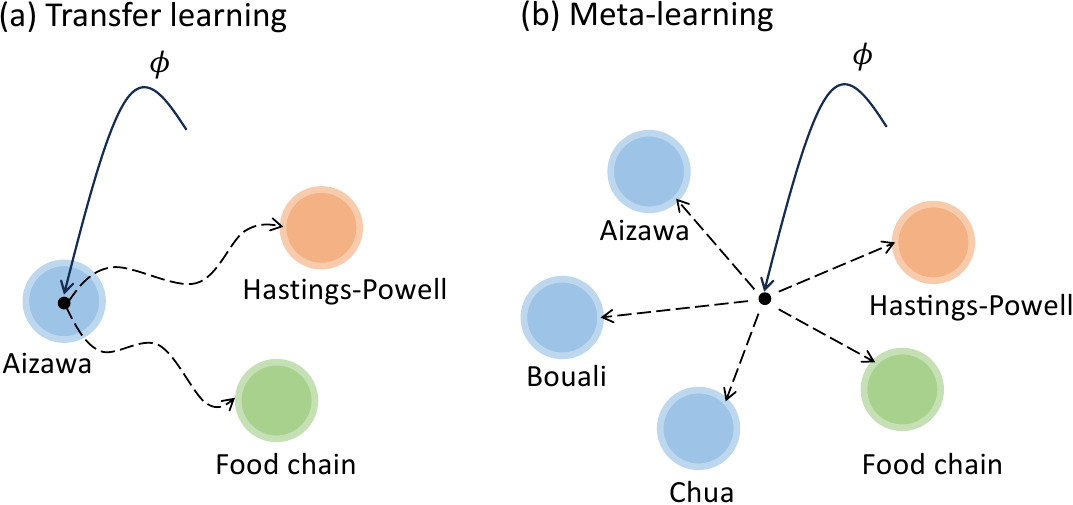} 
\caption{Illustration of transfer learning versus meta-learning. The solid lines denote 
the learning of the initial parameters, and the dashed lines show the path of 
fine-tuning. Blue circles represent the adaptation tasks and other colored circles 
indicate the deployment tasks.}
\label{fig:transfer_meta}
\end{figure}

\begin{figure} [ht!]
\centering
\includegraphics[width=\linewidth]{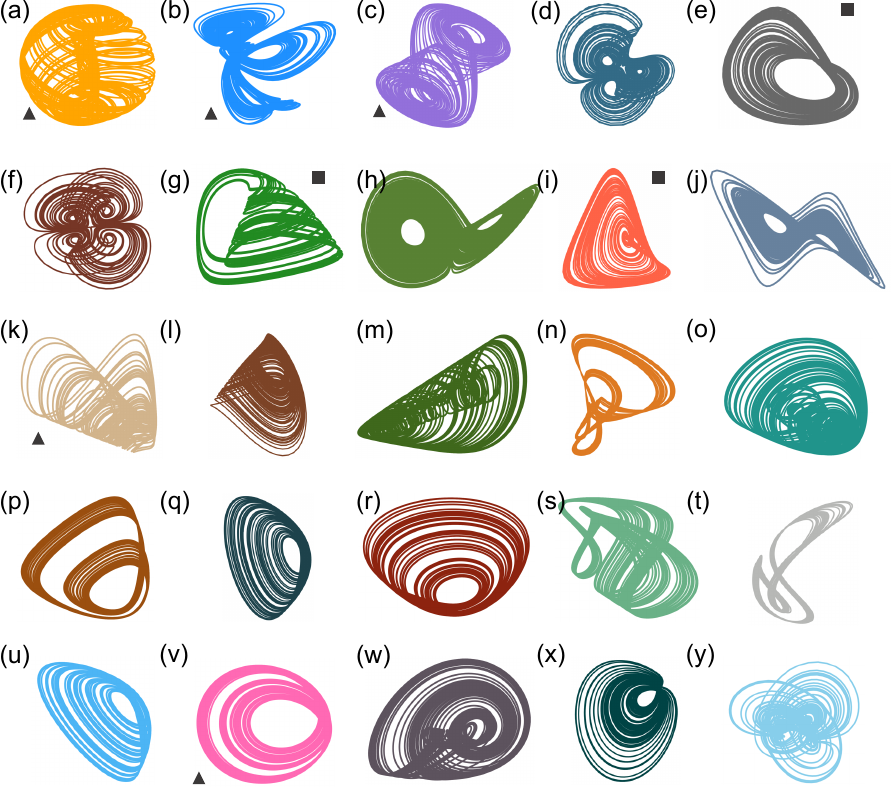} 
\caption{Illustration of the synthetic chaotic systems for adaptation learning during 
the training and the target ecosystems. (a) Aizawa, (b) Bouali, (c) Chua, (d) Dadras, 
(e) Food chain, (f) Four wing, (g) Hastings-Powell, (h) Lorenz, (i) Lotka-Volterra, 
(j) Rikitake, (k-y) Sprott systems. The systems chosen by the greedy algorithm in the 
adaptation phase are marked by triangles and the target ecosystems are marked by 
rectangles.}
\label{fig:chaotic_systems}
\end{figure}

\begin{figure} [ht!]
\centering
\includegraphics[width=\linewidth]{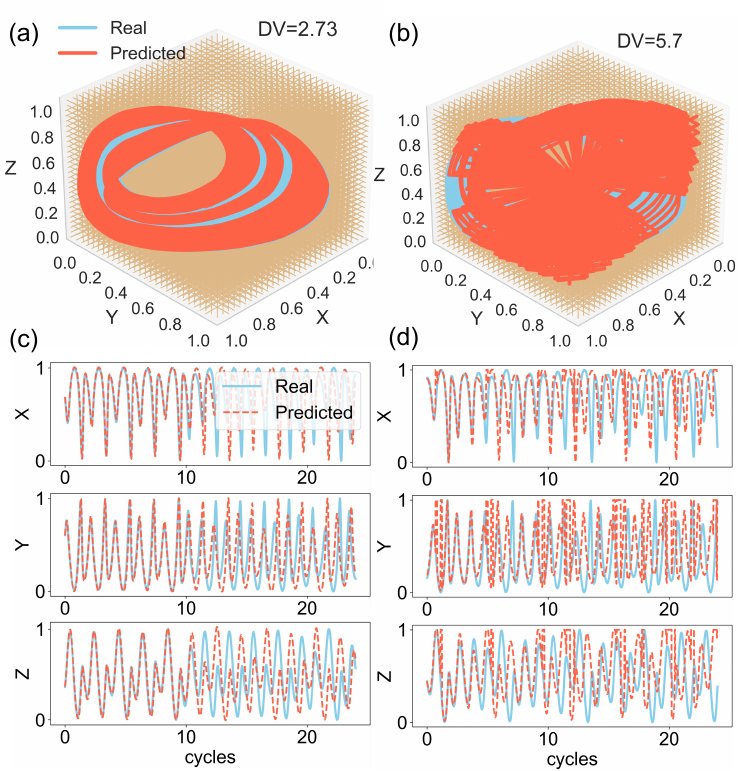} 
\caption{Long-term prediction of the chaotic three-species food chain system by 
meta-learning and standard frameworks. (a,b) Reconstructed attractor by the two 
frameworks. (c,d) Intercepted snippets of the ground truth and machine-learning 
prediction. The results in the left and right columns are generated by the 
meta-learning and the standard frameworks, respectively.}
\label{fig:dv_foodchain}
\end{figure}

It is worth discussing the difference between transfer learning and meta-learning.
Transfer learning, often referred to as fine-tuning, is a technique where a neural 
network trained for one task is applied to another related task. It involves 
leveraging the pre-trained layers of the network, where some layers are ``frozen'' 
(i.e., their weights are not updated), while others are adjusted or added to tailor 
the network to the new task. This approach capitalizes on the existing knowledge of 
the network, particularly its ability to recognize and utilize data patterns. It is 
especially effective when the tasks share common features or structures. Meta-learning, 
or ``learning to learn,'' is at a higher level. It focuses on understanding and 
optimizing the learning process itself rather than acquiring knowledge from the 
data directly. Meta-learning algorithms accumulate experiences from various 
learning tasks and develop strategies to efficiently optimize the parameters 
of a neural network. This process enables the network to quickly adapt to new tasks, 
learning not just specific data patterns but the underlying principles of learning and 
optimization.

In general, in meta-learning, knowledge is learned but not directly (e.g., through 
hyperparameters), where learning can be related to optimization~\cite{gu2018meta}. In 
fact, meta-learning optimizes the hyperparameters of the networks that were not trained, 
whereas transfer learning takes a well-trained network and only adjusts some parts, 
given a new task. Empirically, transfer learning requires the trained and adapted 
tasks to be more similar. Figure~\ref{fig:transfer_meta} presents a comparison between 
transfer learning and meta-learning, where transfer learning trains a neural network 
specifically for the source task (Aizawa system), and fine-tunes the system to each 
target task (e.g., the Hastings-Powell and food-chain systems). In comparison, 
meta-learning trains the neural network to adapt and fine-tune both source tasks and 
target tasks, where the learned knowledge is not directly reflected in the neural 
network parameters but is embedded in the parameter optimization procedure.

\section{Details of the synthetic chaotic systems for adaptation and target ecosystems} \label{appendix_D}

The synthetic chaotic systems employed during the adaptation phase of the training 
and the target ecosystem are illustrated in Fig.~\ref{fig:chaotic_systems}. The system 
equations and parameters used to simulate the time series for these systems are 
listed in Tabs.~\ref{tab:chaos} and \ref{tab:sprott}. 

\begin{table*}[ht!]
\caption{\label{tab:chaos}
Synthetic chaotic systems and target ecosystems}
\begin{ruledtabular}
\begin{tabular}{ccc}
Systems & Equations & Parameters\\
\hline
\\
Aizawa & $\dot{x}=(z-b)x-dy$ & $a=0.95, b=0.7, c=0.6$ \\
& $\dot{y}=dx+(z-b)y$ & $d=3.5, e=0.25, f=0.1$ \\
& $\dot{z}=c+az-z^3/3-(x^2+y^2)(1+e^z)+fzx^3$ \\
\\
Bouali~\cite{bouali2012novel} & $\dot{x}=x(a-y)+\alpha z$ & $\alpha=0.3, \beta=0.05$ \\
& $\dot{y}=-y(b-x^2)$ & $a=4, b=1, c=1.5, s=1$ \\
& $\dot{z}=-x(c-sz)-\beta z$\\
\\
Chua~\cite{matsumoto1984chaotic} & $\dot{x}=\alpha(y-x-ht)$ & $\alpha=15.6, \gamma=1, \beta=28$\\
& $\dot{y}=\gamma(x-y+z)$ & $\mu_0=-1.143, \mu_1=-0.714$\\
& $\dot{z}=-\beta y$ & $ht=\mu_1 x + 0.5(\mu_0-\mu_1)(|x+1|-|x-1|)$\\
\\
Dadras~\cite{dadras2009novel} & $\dot{x}=y-ax+byz$ & $a=3, b=2.7$\\
& $\dot{y}=cy-xz+z$ & $c=1.7, d=2, e=9$\\
& $\dot{z}=dxy-ez$  \\
\\
Four\: wing~\cite{wang20093} & $\dot{x}=ax+yz$ & $a=0.2, b=0.01, c=-0.4$\\
& $\dot{y}=bx+cy-xz$ \\
& $\dot{z}=-z-xy$ \\
\\
Lorenz~\cite{lorenz1963deterministic} & $\dot{x}=\sigma (y-x)$ & $\sigma=10, \rho=28, \beta=2.67$\\
& $\dot{y}=x (\rho-z) - y$ & \\
& $\dot{z}=xy - \beta z$ & \\
\\
Rikitake~\cite{llibre2009global} & $\dot{x}=-\mu x + zy$ & $\mu=2, a=5$\\
& $\dot{y}=-\mu y + x(z-a)$ & \\
& $\dot{z}=1-xy$ & \\
\\
Rossler~\cite{rossler1976equation} & $\dot{x}=-(y+z)$ & $a=0.2, b=0.2, c=5.7$\\
& $\dot{y}=x+ay$ & \\
& $\dot{z}=b+z(x-c)$ & \\
\\
\hline
\\
Food chain & $\dot{R}=R(1-R/{\rm{K}}) - {\rm{x_c y_c}} CR/(R+{\rm{R_0}})$ & $\rm{K=0.98, y_c=2.009, y_p=2.876}$\\
& $\dot{C}={\rm{x_c}} C ({\rm{y_c}} R / (R+{\rm{R_0}})-1) - {\rm{x_p y_p}} PC / (C+{\rm{C_0}})$ & $\rm{x_c=0.4, x_p=0.08}$\\
& $\dot{P}={\rm{x_p}} P ({\rm{y_p}} C / (C+{\rm{C_0}})-1)$ & $\rm{R_0=0.16129, C_0=0.5}$ \\
\\
Hastings-Powell & $\dot{V}=V(1-V)-a_1 VH/ (b_1 V + 1)$ & $a_1=5, a_2=0.1$ \\
& $\dot{H}=a_1VH/(b_1 V + 1) - a_2 H P/ (b_2 H + 1) - d_1 H$ & $b_1=3, b_2=2$ \\
& $\dot{P}=a_2 H P/ (b_2 H + 1) - d_2 P$ & $d_1=0.4, d_2=0.01$ \\
\\
Lotka-Volterra & $\dot{P_i} = r_i P_i (1 - \sum_{j=1}^N a_{ij} P_j)$ & $r_i = [1; 0.72; 1.53; 1.27]$\\
& & $a_{ij}=\begin{pmatrix}
1 & 1.09 & 1.52 & 0 \\
0 & 1 & 0.44 & 1.36 \\
2.33 & 0 & 1 & 0.47 \\
1.21 & 0.51 & 0.35 & 1
\end{pmatrix}$
\end{tabular}
\end{ruledtabular}
\end{table*}

\begin{table}[ht!]
\caption{\label{tab:sprott}
Chaotic sprott systems~\cite{sprott1994some}}
\begin{ruledtabular}
\begin{tabular}{ccc}
Case & Equations &\\
\hline 
\\
0 & $\dot{x}=y$, $\; \dot{y}=-x + yz$, $\; \dot{z}=1-y^2$ \\
\\
1 & $\dot{x}=yz$, $\;\dot{y}=x-y$, $\;\dot{z}=1-xy$ \\
\\
2 & $\dot{x}=yz$, $\;\dot{y}=x-y$, $\;\dot{z}=1-x^2$ \\
\\
3 & $\dot{x}=-y$, $\;\dot{y}=x+z$, $\;\dot{z}=xz+3y^2$ \\
\\
4 & $\dot{x}=yz$, $\;\dot{y}=x^2-y$, $\;\dot{z}=1-4x$ \\
\\
5 & $\dot{x}=y+z$, $\;\dot{y}=-x+0.5y$, $\;\dot{z}=x^2-z$ \\
\\
6 & $\dot{x}=0.4x + z$, $\;\dot{z}=xz-y$, $\;\dot{y}=-x+y$ \\
\\
7 & $\dot{x}=-y + z^2$, $\;\dot{y}=x + 0.5y$, $\;\dot{z}=xz$ \\
\\
8 & $\dot{x}=-0.2y$, $\;\dot{y}=x + z$, $\;\dot{z}=x + y^2 - z$ \\
\\
9 & $\dot{x}=2z$, $\;\dot{y}=-2y + z$, $\;\dot{z}=-x+y+y^2$ \\
\\
10 & $\dot{x}=xy - z$, $\;\dot{y}=x - y$, $\;\dot{z}=x + 0.3z$ \\
\\
11 & $\dot{x}=y + 3.9z$, $\;\dot{y}=0.9x^2 - y$, $\;\dot{z}=1 - x$ \\
\\
12 & $\dot{x}=-z$, $\;\dot{y}=-x^2 - y$, $\;\dot{z}=1.7 + 1.7x + y$ \\
\\
13 & $\dot{x}=-2y$, $\;\dot{y}=x + z^2$, $\;\dot{z}=1 + y - 2z$ \\
\\
14 & $\dot{x}=y$, $\;\dot{y}=x - z$, $\;\dot{z}=x + xz + 2.7y$ \\
\\
15 & $\dot{x}=2.7y + z$, $\;\dot{y}=-x + y^2$, $\;\dot{z}=x + y$ \\
\\
16 & $\dot{x}=-z$, $\;\dot{y}=x - y$, $\;\dot{z}=3.1x + y^2 + 0.5z$ \\
\\
17 & $\dot{x}=0.9 - y$, $\;\dot{y}=0.4 + z$, $\;\dot{z}=xy - z$ \\
\\
18 & $\dot{x}=-x - 4y$, $\;\dot{y}=x + z^2$, $\;\dot{z}=1 + x$ \\
\end{tabular}
\end{ruledtabular}
\end{table}

\section{Results of long-term prediction for two ecosystems}  \label{appendix_E}

\begin{figure} [ht!]
\centering
\includegraphics[width=\linewidth]{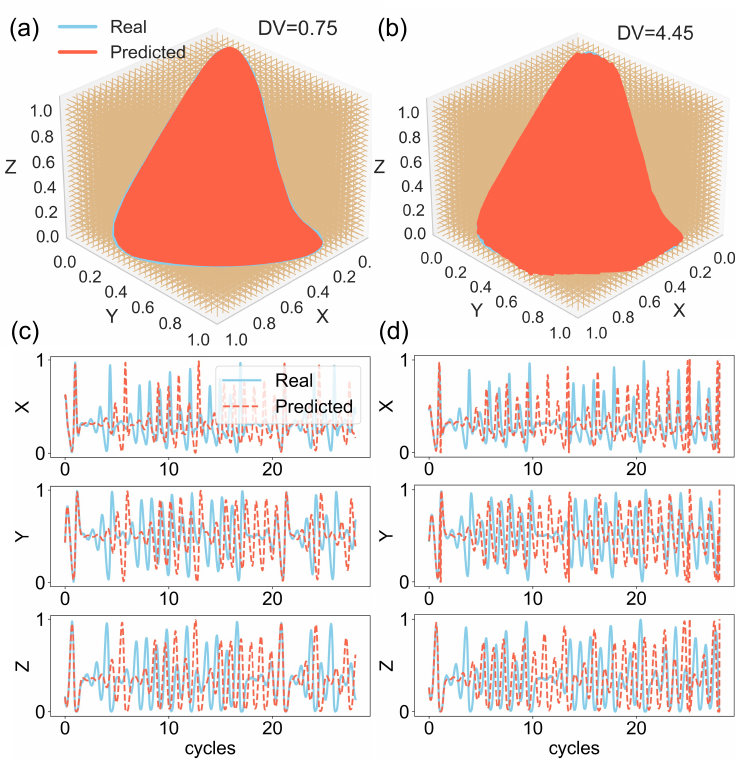} 
\caption{Long-term prediction of the chaotic Lotka-Volterra system by 
meta-learning and standard frameworks. (a,b) Reconstructed attractor by the two 
frameworks. (c,d) Intercepted snippets of the ground truth and machine-learning 
prediction. The results in the left and right columns are generated by the 
meta-learning and the standard frameworks, respectively.}
\label{fig:dv_lotka}
\end{figure}

The three-species food chain system is described by the following set of nonlinear 
differential equations:
\begin{align}
\frac{dR}{dt} &= R(1-\frac{R}{{\rm K}}) - \frac{{\rm{x_c y_c}} CR}{R+{\rm{R_0}}}, \nonumber \\
\frac{dC}{dt} &= {\rm{x_c}} C( \frac{{\rm{y_c}} R}{R+{\rm{R_0}}} - 1) - \frac{{\rm{x_p y_p}} PC}{C+{\rm{C_0}}}, \\
\frac{dP}{dt} &= {\rm{x_p}} P ( \frac{{\rm{y_p}} C}{C+{\rm{C_0}}} - 1 ), \nonumber
\end{align}
where $R$, $C$, and $P$ are the population densities of the resource, consumer, and 
predator species, respectively. The constant parameters 
$\rm K,~ x_c,~ y_c,~ x_p,~ y_p,~ R_0,~ C_0$ are set to be 
$0.98,~ 0.4,~ 2.009,~ 0.08,~ 2.876,~ 0.16129,~ 0.5$, respectively, as stipulated by
a bioenergetics argument~\cite{mccann1994nonlinear}. The statistical results have 
been shown in the second row of Figs.~\ref{fig:dv_comparison}(e) and 
\ref{fig:dv_comparison}(f) in the main text, which compare the DV and prediction 
stability between the meta-learning and standard frameworks. Across all training cycles, 
the meta-learning framework consistently demonstrates a lower testing DV and higher 
prediction stability than the standard model. Figure~\ref{fig:dv_foodchain} exemplifies 
the results of reconstructing the attractor, where the left and right columns are 
generated by meta-learning and standard frameworks, respectively. It can be seen that
meta-learning predicted attractor possesses a more similar ``climate'' to the ground 
truth.

The Lotka-Volterra system is described by 
\begin{align}
\frac{dP_i}{d} = r_i P_i (1 - \sum_{j=1}^N a_{ij} P_j),
\end{align}
where $r_i$ is the growth rate of species $i$ and $a_{ij}$ describes the extent to 
which species $j$ competes with $i$ for resources. We use the four-species competitive 
Lotka-Volterra model defined by~\cite{vano2006chaos}, where the parameters are set as
\begin{align}
r_i = \begin{bmatrix}
       1  \\[0.3em]
       0.72 \\[0.3em]
       1.53 \\[0.3em]
       1.27
     \end{bmatrix},
\quad 
a_{ij} = \begin{bmatrix}
       1 & 1.09 & 1.52 & 0    \\[0.3em]
       0 & 1    & 0.44 & 1.36 \\[0.3em]
       2.33 & 0 & 1    & 0.47 \\[0.3em]
       1.21 & 0.51 & 0.35 & 1
     \end{bmatrix},
\end{align}
The comparative statistical results have been shown in the last row of 
Figs.~\ref{fig:dv_comparison}(e) and \ref{fig:dv_comparison}(f) in the main text.
Similar to the results from the chaotic food-chain system, the meta-learning framework 
consistently demonstrates a lower testing DV, and higher prediction stability than
the standard framework. Figure~\ref{fig:dv_lotka} shows an example of attractor 
reconstruction, where the left and right columns are generated by meta-learning 
and standard frameworks, respectively. Meta-learning generates an attractor that has 
a more similar ``climate'' to the ground truth.

\section{Hyperparameter optimization} \label{appendix_F}

\begin{table}[ht!]
\caption{\label{tab:hyper}
Optimal hyperparameter values}
\begin{ruledtabular}
\begin{tabular}{cc}
Hyperparameters  & Descriptions \\
\hline 
\\
$I_o$=30 & Number of meta-learning outer iterations \\
$I_i$=20 & Number of meta-learning inner iterations \\
$\epsilon$=1.0 & Meta-learning learning rate \\
$\alpha$=$1\times10^{-3}$ & Neural network base learning rate \\
$\beta$=0.9 & Gradient descent momentum term\\
$n_t$=2 & Task sampling number
\\
\end{tabular}
\end{ruledtabular}
\end{table}

In our meta-learning based time-delayed FNN framework, the optimal hyperparameter values 
are determined through Bayesian optimization, which are listed in Table~\ref{tab:hyper}.
To appreciate The role of the optimal hyperparameters, we examine the effects of varying 
the hyperparameters on the prediction performance.

\begin{figure} [ht!]
\centering
\includegraphics[width=\linewidth]{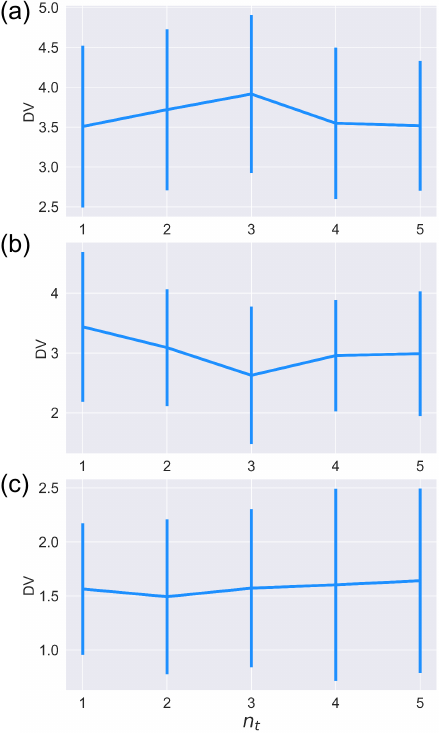} 
\caption{Effect of task sampling number $n_t$ on long-term prediction performance. 
(a-c) DV versus $n_t$ for the Hastings-Powell, food-chain, and Lotka-Volterra systems, 
respectively. Each data point and the associated error bar are obtained from 50 
independent runs.}
\label{fig:sample_number}
\end{figure}

\begin{figure} [ht!]
\centering
\includegraphics[width=\linewidth]{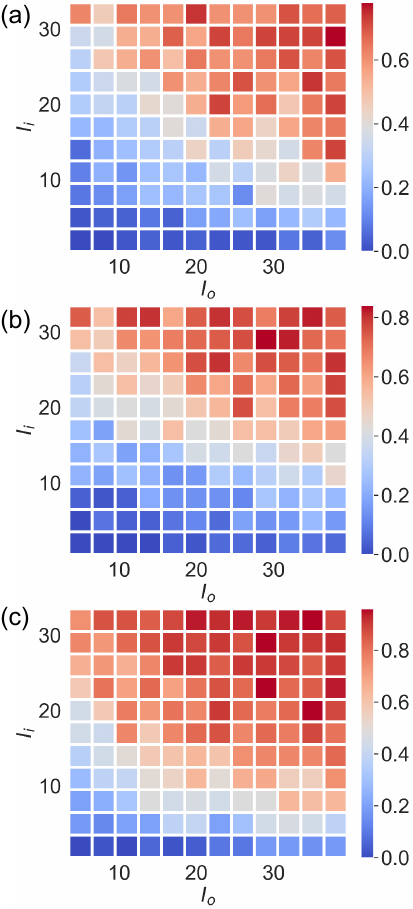} 
\caption{Effects of hyperparameters $I_o$ and $I_i$ on the prediction performance. 
(a-c) Color-coded DV values in the plane of the two hyperparameters for the 
Hastings-Powell, food-chain, and Lotka-Volterra systems, respectively. Each DV value 
is obtained using 50 statistical realizations.}
\label{fig:hyper_iterations}
\end{figure}

\begin{figure} [ht!]
\centering
\includegraphics[width=\linewidth]{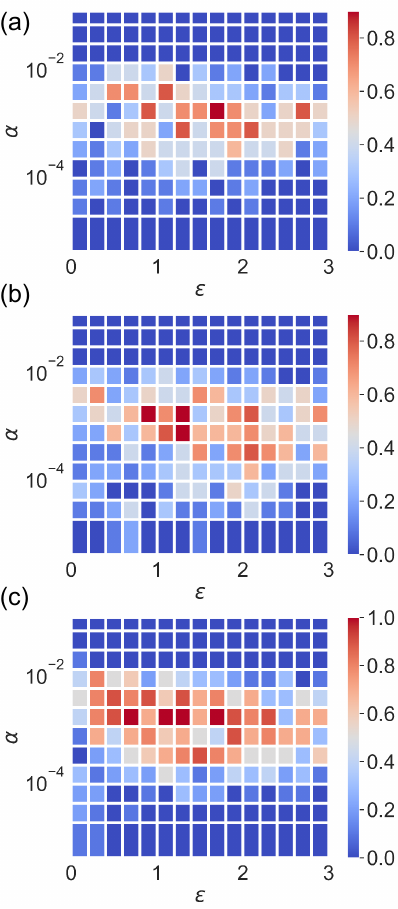} 
\caption{Effects of hyperparameters $\epsilon$ and $\alpha$ on the prediction 
performance. (a-c) Color-coded DV values in the plane of the two hyperparameters 
for the Hastings-Powell, food-chain, and Lotka-Volterra systems, respectively. Each 
DV value is obtained using 50 statistical realizations.}
\label{fig:hyper_iter_lr}
\end{figure}

Figure~\ref{fig:sample_number} shows the influence of the task sampling number $n_t$ 
that determines the number of tasks selected in each meta-learning iteration. Considering
the tradeoff between high performance and computational efficiency, we select $n_t=2$.
Figure~\ref{fig:hyper_iterations} demonstrates the effect $I_o$ and $I_i$, the numbers of 
meta-learning outer and inner iterations, respectively, on the performance of predicting
the long-term dynamics of the ecosystems, where the color-coded values of the 
$R_s({\rm DV}_c)$ are displayed in the two-dimensional parameter plane ($I_o,I_i$). 
Specifically, $I_o$ is the number of times of meta-learning updates in the adaptation 
phase and $I_i$ is the epoch of the neural network to learn each task. As $I_o$ and $I_i$
increase, the performance is dramatically improved. For high performance and low 
computational time, we select $I_o=30$ and $I_i=20$.

Figure~\ref{fig:hyper_iter_lr} demonstrates the effect of meta-learning learning rate 
$\epsilon$ and neural network base learning rate $\alpha$ on the performance of predicting
the attractors of the ecosystems, where the color-coded values of the $R_s({\rm DV}_c)$ 
are displayed in the two-dimensional parameter plane ($\epsilon,\alpha$). Particularly,
the learning rate controls the step size of each outer iteration taken, and the neural 
network base learning rate controls the learning step size of the inner iterations. It
can be seen that the prediction performance depends more on $\alpha$ than on $\epsilon$. 
 we select $\epsilon=1.0$ and $\alpha=1\times 10^{-3}$ in our work.

\section{Short-term prediction} \label{appendix_G}

\begin{figure} [ht!]
\centering
\includegraphics[width=\linewidth]{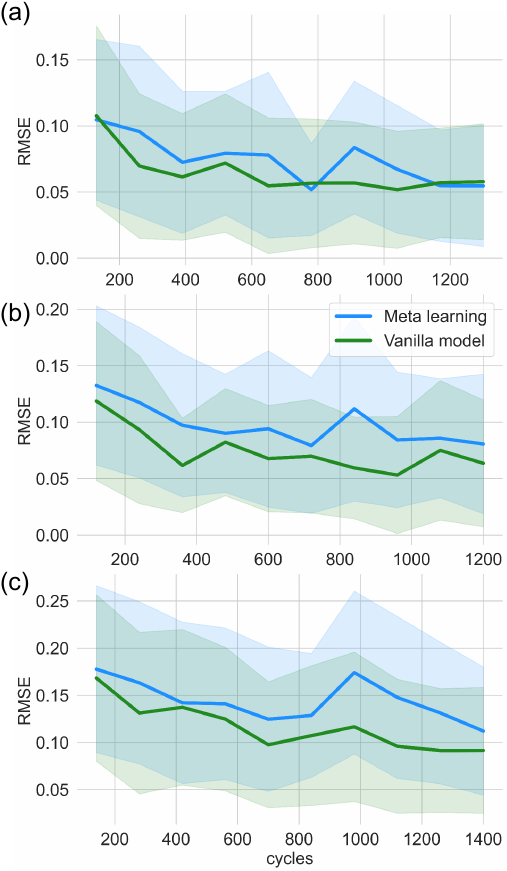} 
\caption{Short-term prediction of ecosystems by meta-learning and standard frameworks. 
(a-c) RMSE versus the training length for the Hastings-Powell, chaotic food-chain, and 
chaotic Lotka-Volterra systems, respectively. The RMSEs and variabilities (in shade) 
are calculated from the ensemble of 50 independently trained machines.}
\label{fig:short_rmse}
\end{figure}

While the primary focus of our study is on reducing the training time for 
long-term prediction, we have also studied the short-term prediction for both the 
meta-learning and standard frameworks. Here, by ``short-term'' we mean a few cycles
of oscillations of the target ecosystem. To be concrete, for the Hastings-Powell, 
food-chain, and Lotka-Volterra systems, we use 300, 400, and 360 time steps, 
corresponding approximately to 3.9, 4.8, and 5.1 cycles of oscillation, respectively. 
To characterize the performance of short-term prediction, we use the root-mean-square 
error (RMSE) defined as
\begin{align}
{\rm RMSE(\bold y, \hat{\bold y})} = \sqrt{\frac{1}{T_p} \frac{1}{N} [y_n(t) - \hat{y}_n(t)]^2},
\end{align}
where $\bold y$ and $\hat{\bold y}$ are the true and predicted time series, 
respectively, $N=3$ is the system dimension, and $T_p$ denotes the prediction time.
Figure~\ref{fig:short_rmse} presents results for the three ecosystems. It can be seen
that meta-learning and standard frameworks exhibit similar performance, suggesting
that, with respect to short-term prediction of the state evolution, meta-learning
offers no apparent advantage. A plausible reason is that meta-learning requires an 
adaptation phase during which learning is conducted based on time series from a number
of synthetic chaotic systems. While exhibiting similar attractors, the detailed 
dynamical evolution of these systems are quite different from that of the target 
ecosystem, compromising the accuracy of short-term prediction.

% \bibliographystyle{agsm}
%\bibliography{COC}

\end{document}